# A Multipath Energy-Conserving Routing Protocol for Lifetime Improvement of Wireless Ad Hoc Networks

Omar Smail[1]*, Bernard Cousin[2], Zoulikha Mekkakia[1] and Rachida Mekki[1]

**Abstract**
Ad hoc networks are wireless mobile networks that can operate without infrastructure and without centralized network management. Traditional techniques of routing are not well adapted. Indeed, their lack of reactivity with respect to the variability of network changes makes them difficult to use. Moreover, conserving energy is a critical concern in the design of routing protocols for ad hoc networks, because most mobile nodes operate with limited battery capacity, and the energy depletion of a node affects not only the node itself but also the overall network lifetime. In all proposed single-path routing schemes a new path-discovery process is required once a path failure is detected, and this process causes delay and wastage of node resources. A multipath routing scheme is an alternative to maximize the network lifetime. In this paper, we propose an energy-efficient multipath routing protocol, called AOMR-LM (Ad hoc On-demand Multipath Routing with Lifetime Maximization), which preserves the residual energy of nodes and balances the consumed energy to increase the network lifetime. To achieve this goal, we used the residual energy of nodes for calculating the node energy level. The multipath selection mechanism uses this energy level to classify the paths. Two parameters are analyzed: the energy threshold $\beta$ and the coefficient $\alpha$. These parameters are required to classify the nodes and to ensure the preservation of node energy. Our protocol improves the performance of mobile ad hoc networks by prolonging the lifetime of the network. This novel protocol has been compared with other protocols: AOMDV and ZD-AOMDV. The protocol performance has been evaluated in terms of network lifetime, energy consumption, and end-to-end delay.

**Keywords**: Ad hoc network, Multipath routing, Energy efficiency, Network lifetime.

## 1 Introduction

An ad hoc network is characterized by frequent changes in the network topology, limited bandwidth availability, and limited power of nodes. The ad hoc network topology changes frequently as nodes are able to move collectively or individually and often in an unpredictable way. When one node moves out of or into the transmission range of another node, the wireless link between the two becomes down or up. Another cause of the topological changes is the instability of the wireless link quality, which might become high or low due to the signal fading (obstacles between the two wireless nodes), interference from other signals, or the change in the node transmission power level. In the following we refer to the disappearance of the wireless link for any reason as link failure. Mobile nodes are battery powered; when they run out of battery power, they fail. All these types of failures increase the intensity of changes in the network topology. As the nodes have a limited communication range, the path from source to destination usually has multiple hops (the data packets are retransmitted by some intermediate nodes); hence these characteristics make route discovery complex.

The routing problem in mobile wireless networks has attracted considerable interest in the research community. Several research studies have focused on routing protocols of ad hoc networks [1–5]. While the proposed protocols have some relevant characteristics, they have limitations, especially in the case of high mobility of nodes or high network load. The traditional approach of routing in mobile ad hoc networks adopts a single active route between a source node and a destination node for a given communication. This usually uses proactive [6,7] or reactive (on-demand) [8,9] routing protocols. In [10], it is shown that proactive protocols are very expensive in terms of energy consumption compared to reactive protocols, because of the large routing overhead incurred in the former. But reactive protocols suffer from high latency during the process of discovering fresh paths, especially in large networks and dense networks [11].

*Correspondence: omar.smail@univ-usto.dz
[1]Faculty of Mathematics and Informatics. Computer Science Department. University of Sciences and Technology, 'Mohamed Boudiaf' USTO-MB, Oran, Algeria.
[2]IRISA/University of Rennes 1, France.

In recent years, the research community has focused on the improvement of ad hoc routing, with the development of several routing mechanisms. Multipath routing seems to be an effective mechanism in ad hoc networks with high mobility and high load to guard against the problem of frequent changes of the network topology caused mainly by link failures. The concept of multipath routing is that the source node is given the choice between multiple paths to reach a given destination. The multiple paths can be used alternately; the data traffic takes a single path at one time or several paths simultaneously. A multipath between a source and destination must be chosen wisely so that a path failure does not disturb other paths as less as possible. There are two types of disjoint paths: link disjoint paths and node disjoint paths. Node disjoint paths have no node in common except the source and destination. Link disjoint paths have no common links, while they may share some nodes. Any path of a multipath can be used to transmit a data packet between a source and a destination. Thus to maximize the data flow and to get a larger share of the network bandwidth, the data packets of a flow between a source and a destination can be split between the paths [12]. Multipath routing is highly suitable for multimedia applications, to ensure secure transmission and it is proposed for industrial ad hoc networks for improving reliability and determinacy of data transmission. Another benefit of multipath routing protocols is the reduction of the routing overhead, for which several multipath routing protocols have been developed [8,13–15]; these works propose a single path discovery process able to build several link or node disjoint routes towards the destination based on broadcast requests. The approaches in [16,17] are improvements of single path routing protocols. They contribute in reducing delays and increasing throughput because they resist node mobility in comparison with single path approaches.

In ad hoc networks, each node has a power battery and a limited energy supply. Over time, nodes deplete their energy supplies and are eventually removed from the network, which constrains the packet routing. Some exhausted nodes may be critical for packet transmission when they are involved in the only path from the source to the destination node. To solve this inefficiency, we propose a protocol which extends the lifetime of the network. We used the energy levels of nodes to classify the nodes of paths between a source and a destination. The node energy level is calculated based its residual energy and on nodes energy that contributes in the routing; this information is collected in the discovery of multiple paths using the same principle as in the discovery of a single path. Two parameters are introduced, the energy threshold $\beta$ and the coefficient $\alpha$, to define the node class in order to preserve the consumed energy. Our protocol selects the nodes with similar classes to construct a given path. At the end of the route discovery process the source node will have a set of homogeneous paths in terms of energy. The best classified paths are selected to balance the energy consumed between the different paths of a multipath. The proposed technique discards nodes with a critical energy level and makes sure they do not participate in routing, otherwise we may have several links failures caused by the depletion of nodes.

The paper is organized as follows. Section 2 provides a review of related works for energy-aware routing protocols in wireless ad hoc networks. Section 3 presents an overview of the AOMDV protocol and motivates its exploitation. Section 4 gives the design details of our AOMR-LM protocol. Section 5 provides the simulation results of its performance evaluation. Section 6 concludes this paper.

## 2 Energy-Aware Routing Protocols

The main objective of ad hoc routing protocols is to deliver data packets among network nodes efficiently, which depends on the node lifetime. The lifetime of a node is directly proportional to its battery's energy. The node battery's energy is primarily consumed while transmitting and receiving packets or computing. As ad hoc networks are multihop networks, one node may be involved in retransmission of packets sent between a source node and a destination node. The nodes participating in packet transmission, can exhausted their energies and removed from the network. This affects the reliability of the packet delivery service of the network. So in a multihop communication, the selection of the appropriate nodes by the routing protocol is very important. Thus, routing algorithms play an important role in saving the communication energy, this extends the life of the nodes and thus of the whole network. Many works focusing on energy-aware routing protocols exist in the literature [18-23]. These energy-aware routing protocols can be single path or multipath. In the following, the main proposed works about energy-aware single path routing protocols are described.

In [24] the authors presented several routing protocols with energy measurements that track energy efficiently. Minimum Total Transmission Power Routing (MTPR) [25] was initially developed to minimize the power consumption of nodes involved in a path. However, such protocols do not take node energy capacity into account. Thus, the energy consumption is not fairly shared among nodes in the network. Min-Max Battery Cost Routing (MMBCR) [26] considers the remaining power of nodes as a metric for the acquisition of paths in order to extend the lifetime of the network. But, this protocol cannot guarantee that the energy consumption will be minimized. The main advantage of on-demand routing comes from the reduction of the routing overhead, as a high routing overhead has a significant impact on the performance of wireless networks. An on-demand routing protocol attempts to discover a route toward a destination when a data packet is presented to be transmitted to that destination. Providing multiple paths for transmitting the data packets is beneficial in ad hoc network communications: it can be useful for improving the effective bandwidth of communication, reducing the routing overhead, and decreasing delivery delays. Recently, several on-demand routing multipath protocols have been proposed that preserve energy in order to avoid network failures as long as possible.

In [27], Liu et al. propose a new algorithm called Collision-Constrained minimum Energy Node disjoint multipath routing Algorithm for ad hoc networks (ECCA). The algorithm is a trade-off between collision avoidance and energy saving. It calculates an upper limit of the correlation factor depending on the service required and finds a disjoint multipath with energy which satisfies the upper limit. ECCA can significantly reduce the packet loss rate and the consumed energy. Hence, it attempts to provide energy savings at the nodes. But it does not take into account the power status of the nodes and thus the duration of availability of the network. The Multipath Energy-Efficient Routing protocol (MEER) [28] uses a control mechanism for rational power. The route discovery process in which the source tries to discover routes with high energy is based on SMR (Split Multipath Routing) [29]. This protocol protects the nodes from consuming too much energy compared to the other nodes in the network. However the proposed approach is not based on disjoint routes, and thus it cannot exploit the benefits of path disjointness, which is useful to balance the energy consumption among network nodes. Max-Min Residual Energy (MMRE-AOMDV) [30] is a multipath routing protocol based on AOMDV (Ad hoc On-demand Multipath Distance Vector) [11]. This protocol finds the minimal nodal residual energy of nodes of each path and then selects the path with maximal residual energy to forward the data packets. The MMRE-AOMDV protocol uses the routing information already available in the underlying AOMDV protocol. Thus little additional overhead is required for the computation of the maximal nodal residual energy in the route. The MMRE-AOMDV protocol has two main components: first the computation of the nodal residual energy of each route during the route discovery process, and second the sorting of routes by the descending order of their nodal residual energy. It uses the route with maximal nodal residual energy to forward data packets. Simulation results showed that the proposed MMRE-AOMDV routing protocol performed better than AOMDV in terms of packet delivery fraction, throughput, and network lifetime. But this protocol does not evaluate the energy consumption and its impact on network performance, knowing that the network life depends on the node expiration, which in turn depends upon energy consumption. In [31], Liu et al. proposed the Multipath Routing protocol for Network Lifetime Maximization (MRNLM), a protocol that defines a threshold to optimize the forwarding mechanism. It proposes an energy-cost function and uses the function as the criterion for multiple path selection. During the transmission phase, they use a method called "data transmission in multiple paths one by one" to balance the energy consumption on the multiple paths. It is shown that MRNLM consumes less energy than AOMDV but does not improve the end-to-end delay. Multipath Multimedia Dynamic Source Routing (MMDSR) [32] is a multipath routing protocol able to self-configure dynamically according to network states. The authors used cross-layer techniques to improve the end-to-end performance of video-streaming services over networks using IEEE 802.11e. MMDSR uses an analytical model to estimate the path error probability. This model is used by the routing scheme to estimate the lifetime of paths. In this way, it is hoped that proper proactive decisions can be taken before the paths are broken. However, the comparison with DSR (Dynamic Source Routing) is not fair since it is not a multipath routing protocol. The Multipath Energy Efficient Routing Protocol (MEERP) [33] is an extension of the existing routing protocol AOMDV. Route discovery is modified in MEERP, whereby each intermediate node is prohibited from generating a route reply message. The proposed protocol selects energy-efficient multiple node disjoint paths based on the residual energy and successful transmission rate. In this algorithm, multiple routing paths are selected. However, only one path is used for data transmission at a given time. During the path-discovery process, each intermediate node calculates the cumulative node cost until the destination, and the path with the highest cost is selected. This cost depends upon two measurements: the successful transmission rate and the residual energy of the node. The simulation results show that the proposed scheme can achieve a great improvement of the network lifetime by reducing end-to-end delay and overhead. In [34], the authors propose a multipath routing protocol based on AODV routing algorithm (ZD-AOMDV). The represented protocol in this study tries to discover the distinct paths between source and destination nodes with using Omni directional antennas, to send information through these simultaneously. This protocol counts the number of active neighbors for each path, and finally it chooses some paths for sending information in which each node has lower number of active neighbors all together. Here, active neighbors of a node are defined as nodes that have previously received the RREQ (Route Request).The aim of this work is to try to improve the energy efficiency of ad hoc networks.

All of the above studies solve the problem of energy conservation, but the majority of power-saving mechanisms are based only on the remaining power cannot be used to establish the best route between source and destination nodes. On one hand, if a node is willing to accept all route requests only because it currently has enough residual battery capacity, too much traffic load may be routed through that node. On the other hand, excessive energy savings neglects the power consumption at the individual nodes, it results network partitioning due to nodes battery exhaustion. Indeed, it reduces network performance. Hence, shared and balanced energy consumption is a remedy for those types of problems. Finally, the majority of these protocols have been compared only with original protocols (AODV, AOMDV, DSR, SMR, ...), which do not explicitly consider energy consumption, and thus these performance evaluations are not fair. We will use ZD-AOMDV [34] as a reference for our performance evaluation because it aims to improve the lifetime of the ad hoc network and has the same characteristics as our protocol, namely its reactivity, multipath character, and use of AOMDV as its basic protocol.

## 3 AOMDV Overview

The existing routing protocols in ad hoc networks such as proactive and reactive protocols can be modified to incorporate a power control function which prolongs the network lifetime and optimizes energy consumption. We chose

AOMDV (Ad hoc On-demand Multipath Distance Vector) [11] as the basis for our protocol. AOMDV has been proven to be a good protocol that uses multipath routes; all of its paths are disjoint paths and it can guarantee a loop-free path because it allows only alternate routes with lower hop counts. The AOMDV protocol performs a route discovery process between the source node and the desired destination node when the source needs to send a data packet and the path to the destination is not known. In this process, route query (RREQ) messages are broadcasted by every node in the network. The destination sends a route reply (RREP) message for all of the received RREQ packets. An intermediate node forwards a received RREP packet to the neighbor that is along the path to the source. This discovery process can be exploited to collect fresh node information, such as residual energy, load level, and so on.

Several studies [35–37] have shown that the AOMDV protocol is more robust and performs better in most of the simulated scenarios. So we selected this protocol, instead of any other reactive protocol (such as SMR [29]), as the reference for performance evaluation of our protocol.

## 4 The Ad hoc On-demand Multipath Routing with Lifetime Maximization protocol

In this section, an improved routing protocol, named Ad hoc On-demand Multipath Routing with Lifetime Maximization (AOMR-LM), is presented. AOMR-LM is a multipath routing protocol based on AOMDV protocol, with a new path classification mechanism according to the energy level of the nodes forming these paths, which can be high, average, or low. The idea is to build homogeneous paths in terms of energy, this can balance the consumption energy of nodes and avoid link failures due to nodes energy depletion. The protocol sets an energy threshold and a coefficient, used to define the class of each node. These parameters are required to forward the reply packet decision. This idea helps at prolonging the network lifetime and improves the energy performance in mobile ad hoc networks. In the single path routing AODV protocol, maintenance of paths occurs by sending periodic short messages (called HELLO message). If three consecutive HELLO messages are not received by a node from a neighboring node, the node considers the link to be broken. When the node receives a data packet which should be routed via this broken link, a route error message (RERR, Route ERRor) is sent back to the packet source indicating the broken link. In this case a new path discovery process needed, which adds an extra cost in terms of delay, throughput, control messages, and considerable energy consumption for the additional transfer of control messages. Among the reasons of failure of a node is the limitation of its battery energy. Due to these node failures, links in a path may become temporarily unavailable and make the path invalid. The routing algorithm decides which of the network nodes need to be selected in a particular path. Selection of nodes without taking into account the node energy when determining the paths leads to an umbalanced energy level in the network. Nodes on the minimum energy paths could quickly become drained while other nodes remain intact. This will result in the early death of some nodes. For this purpose, multipath routing has been shown to be effective since it distributes the traffic load among more nodes and in proportion to their residual energies. When the energy consumption among nodes is more balanced, the mean time to node failure is prolonged and so the network lifetime. The proposed AOMR-LM protocol is a reactive protocol for multipath routing. Our selection mechanism preserves the residual energy of nodes and balances the consumed energy. This prolongs the network lifetime and improves network performance.

In the next sections, we first introduce define some assumptions and then provide the main details of multipath discovery, selection, data transmission, and maintenance procedures.

### 4.1 Problem definition

An ad hoc wireless network is represented by an undirected graph, $G = (V, E)$, where $V$ is the set of network nodes and $E$ is the set of network bidirectional links. Let $w(u)$, $u \in V$, represent the residual energy at node $u$.

Let $c(u, v)$, $(u, v) \in E$, be the energy required to transmit a packet from node $u$ to node $v$. We assume that $c(u, v) = c(v, u)$ for all $(u, v) \in E$. If $P$ is any set of paths of $G$; then we consider $P(u_0, u_n)$ to be the set of paths between nodes $u_0$ and node $u_n$. Let $P_i(u_0, u_n) = u_0, u_{i1}, \ldots, u_{im}, \ldots, u_n$, be the $i^{th}$ path in $P(u_0, u_n)$, the source node is noted by $u_0$ or $u_{i0}$ and the destination node is noted by $u_n$ or $u_{in}$. We define $\overline{P_i(u_0, u_n)}$ as the number of nodes on the path $P_i(u_0, u_n)$ and $\overline{P(u_0, u_n)}$ as the number of paths in the multipath discovery process. The sum of the residual energy of a path $P_i(u_0, u_n)$, denoted by $e_{sum}(P_i(u_0, u_n))$ is given by:

$$e_{sum}(P_i(u_0, u_n)) = \sum_{u_{i0}}^{u_{in}} w(u_{i_j}) \qquad (1)$$

We consider $e_{sum}(P(u_0, u_n))$ to be the sum of the residual energies of all paths belonging to the set $P(u_0, u_n)$:

$$e_{sum}(P(u_0, u_n)) = \sum_{P(u_0, u_n)} e_{sum}(P_i(u_0, u_n)) \qquad (2)$$

Let $e_{average}(P_i(u_0, u_n))$ be the average residual energy of one node belonging to a path, $e_{average}(P_i(u_0, u_n))$ is given by:

$$e_{average}(P_i(u_0, u_n)) = \frac{e_{sum}(P_i(u_0, u_n))}{\overline{P_i(u_0, u_n)}} \qquad (3)$$

The average residual energy of a set of paths between a same pair of nodes is denoted by $e_{averageNet}(P(u_0, u_n))$, which is the average residual energy of nodes that participated in the multipath discovery process between one source node $u_0$ and one destination node $u_n$, $e_{averageNet}(P(u_0, u_n))$ is given by:

$$e_{averageNet}(P(u_0, u_n)) = \frac{e_{sum}(P(u_0, u_n))}{\sum \overline{P_i(u_0, u_n)}} \quad \text{Where } P_i \in P \qquad (4)$$

We define $e_{level}^{u_0,u_n}(u_j)$, the energy level of node $u_j$ during a discovery process between a source node $u_0$ and a destination node $u_n$, given by:

$$e_{level}^{u_0,u_n}(u_j) = \frac{w(u_j)}{e_{averageNet}(P(u_o,u_n))} \qquad (5)$$

Given a source *s*, a destination *d*, and a data packet to be routed, the source initiates the multipath discovery. This discovery provides multiple disjoint paths with energy efficiency properties. The path selection is based on the residual energy level of its nodes to preserve the node energy; and to balance the energy consumed over a large set of paths we choose to classify paths and transmit the data by choosing one path of the available highest class.

### 4.2 Multipath discovery

Based on the on-demand routing scheme, the source node starts the multiple paths discovery process to create a set of paths able to forward data towards the destination node from the source node. When a source requires a route toward a destination, the source checks its routing table for any available path toward this destination. If a path is not present or is invalid, the source performs route discovery: it broadcasts an RREQ (Route Request) message to all of its neighbors. When a node receives an RREQ, it ensures that the received RREQ is not a duplicate RREQ by comparing the RREQs' identifiers, in order to prevent looping paths. Otherwise, the RREQ-receiving nodes verify whether they have any valid path toward the destination in their routing tables. If they have, they forward the RREQ to those valid path neighbors. In this case, for our protocol, returning back a RREP (Route Reply message) to the source is not desirable because we would like to collect fresh energy information. Otherwise, the receiving node retransmits the RREQ message to all of its neighboring nodes to find the paths toward the destination. When the destination receives the first RREQ, it waits for a certain time and collects all other RREQs arriving during this time interval. Our protocol uses the destination sequence number in the same way as AOMDV to indicate the freshness of the route, which ensures loop-freedom. Moreover, we use the notion of an advertised hop count to maintain the multiple paths for the same sequence number. The advertised hop count contains the hop count of the longest path allowed. Once the sequence number changes, the advertised hop count is reset and remains unchanged for this sequence number. Indeed a node builds an alternate path to a certain destination node via a neighboring node only if this alternate path has a smaller advertised hop count.

Several changes are needed in the AOMDV route-discovery procedure to enable computation of the sum and average nodal residual energy of the network. Each RREQ now carries an additional field, called $e_{sum}(P_i(u_0, u_j))$, which represents the sum of residual energy from the source to the current node $u_j$. Another field $w(u)$ is added, so that a node knows the residual energy of its neighbors. When an intermediate node receives an RREQ, it increases the field $e_{sum}(P_i(u_0, u_j))$ by the value of its residual energy.

The multipath discovery must take into account the message sequence number in order to ensure the freshness of paths and the maximum hop count for all the paths, denoted respectively by *seqnum* and *advertised_hopcount*. Any RREQ message received with a sequence number lower than the largest sequence number received in any previous RREQ message from that source towards that destination is discarded. The same process is repeated until the RREQ message reaches its final destination; see Algorithm 1.

```
if (seqnum^d_i < seqnum^d_j) then seqnum^d_i := seqnum^d_j;
    if (i ≠ d ) then
        e_sum(P(u_0, u_j)):= e_sum(P(u_0, u_j))+ w(u_i);
        advertised_hopcount^d_i := ∞;
        route_list^d_i := NULL;
        insert ( j , advertised_hopcount^d_j +1,w(u_j) ) into route_list^d_i;
    else
        advertised_hopcount^d_i := 0;
    endif
elseif (seqnum^d_i = seqnum^d_j) and
((advertised_hopcount^d_i,i) > (advertised_hopcount^d_j,j))
    then
        e_sum(P(u_0, u_j)):= e_sum(P(u_0, u_j))+w(u_i);
        insert ( j , advertised_hopcount^d_j +1, w(u_j)) into route_list^d_i;
endif
```

**Algorithm 1: AOMR-LM path discovery process**

Figure 1 shows the structure of an entry of the routing table of a node. For each destination known by the node there is an entry. *Route_list* contains all known neighboring nodes of a node which leads to that destination. Each neighbor for that destination is identified its *nexthop* address, and the *hopcount* field is the number of hops required to reach that destination using this neighbor. We add two new fields, $w(u)$ and *marked_node*, in the *route_list*. The field $w(u)$ denotes the residual energy of a node and the field *marked_node* indicates whether or not a node has been selected by a reverse path (explained later).

| Destination |
|---|
| Sequence_number |
| Advertised_hopcount |
| Route_list |
| {(nexthop1, hopcount1, $w(u_1)$, marked_node), (nexthop2, hopcount2, $w(u_2)$, marked_node), …} |
| Expiration timeout |

**Figure 1: Structure of a routing table entry for AOMR-LM.**

### 4.3 Multipath selection

After reception of the first RREQ packet, the destination node waits for a certain period of time *RREQ_Wait_time* before starting the route selection procedure. This period of waiting generates an additional delay for the multipath routing selection. However the results of the evaluations (Section 5) demonstrates its low impact on end-to-end delay. When this period of time expires, the destination node generates a route reply (RREP) message and sends it back to the source. In the conventional AOMDV, the propagation of route request messages from the source towards the destination establishes multiple paths. However, it does not consider the residual energy of nodes. In our protocol, the energy of a node $w(u_j)$ is considered when the RREP packet is forwarded back. So the selection of the next node toward which the RREP packet is forwarded depends on the node class. To determine the belonging of a node to one of the three classes, we define two thresholds, denoted $\alpha$ and $\beta$. $\alpha$ being the lower threshold which separates the low class from the middle class, $\beta$ the upper threshold which separates the middle class from the upper class. The energy threshold $\beta$ is computed for all network nodes. So for our solution, we assume that the threshold $\beta$ is the average residual energy of the network. If the energy of node $u_j$ (i.e. $w(u_j)$) is above the threshold ($w(u_j) \geq \beta$), the node $u_j$ has a high probability of being able to transmit data packets, and thus it is classified as high class. When $w(u_j) < \beta$, the definition of the node class depends on the threshold $\beta$ and the coefficient $\alpha$, which is necessary for the forward decision. The coefficient $\alpha$ is introduced to determine the node's capacity to support data traffic in terms of energy. This coefficient decides the node participation in routing.

When $w(u_j) < \beta$ then $\dfrac{w(u_j)}{\beta} < 1$. Let $\alpha < 1$; we obtain:

$$\alpha * \frac{w(u_j)}{\beta} < 1 \qquad (6)$$

To analyze the values of the coefficient $\alpha$, we define $T_{Net}(P(u_0,u_n))$, which describes the effective participation of $K$ network nodes in the transfer of data on paths of $P(u_0, u_n)$. The rest of the nodes do not affect the data transfer, so they will not be considered in the formulation of our solution. We have:

$$T_{Net}(P(u_0,u_n)) = \prod_{j=1}^{K} \alpha * \frac{w(u_j)}{\beta} \leq \left(\frac{\alpha}{\beta}\right)^K * \prod_{j=1}^{K} w(u_j)$$

Applying the inequality of arithmetic and geometric means

$$\frac{\alpha_1 x_1 + \alpha_2 x_2 + ... + \alpha_n x_n}{\alpha} \geq \sqrt[\alpha]{x_1^{\alpha_1} x_2^{\alpha_2} ... x_n^{\alpha_n}}$$

where $\alpha = \alpha_1 + \alpha_2 + ... + \alpha_n > 0$,

if we take $n = K, \alpha_i = 1$, and $x_i = w(u_i)$

we obtain $\left(\prod_{j=1}^{K} w(u_j)\right)^{\frac{1}{K}} \leq \left(\frac{1}{K} \sum_{j=1}^{K} w(u_j)\right)$

Then we have

$$\left(\frac{\alpha}{\beta}\right)^K * \prod_{j=1}^{K} w(u_j) \leq \left(\frac{\alpha}{\beta}\right)^K * \left(\frac{1}{K} \sum_{j=1}^{K} w(u_j)\right)^K$$

which proves that

$$T_{Net}(P(u_0,u_n)) = \prod_{j=1}^{K} \alpha * \frac{w(u_j)}{\beta}$$

$$\leq \left(\frac{\alpha}{\beta}\right)^K * \left(\frac{1}{K} \sum_{j=1}^{K} w(u_j)\right)^K$$

To simplify the analysis, we pose $\sum_{j=1}^{K} \dfrac{w(u_j)}{K} \approx e_{averageNet}(P(u_0,u_n))$, since the sum of the residual energy of nodes divided by the number of nodes gives the same value calculated by Equation (4).

It is considered that if the residual energy of a node is greater than the average residual energy of the network, then this node has sufficient energy and has a high probability of transmitting more data packets before being exhausted. This case corresponds to ($w(u_j) \geq \beta$), so:

$$\beta = e_{averageNet}(P(u_0,u_n)) \qquad (7)$$

Therefore: $T_{Net} = \prod_{j=1}^{K} \alpha * \dfrac{w(u_j)}{\beta} \leq \alpha^k$, and we obtain:

$$\alpha \geq (T_{Net})^{\frac{1}{K}} \qquad (8)$$

Following (6), we obtain

$$\alpha < \frac{\beta}{w(u_j)} \qquad (9)$$

With (8) and (9), we have

$$\alpha \in \left[(T_{Net})^{\frac{1}{K}}, \frac{\beta}{w(u_j)}\right] \qquad (10)$$

After receiving an RREP message, an intermediate node checks whether the RREP message has already been received (the intermediate node identifies each RREP message by the IP address of the originating node, and the destination sequence number in order to detect a duplicate message). If the RREP message has already been received (or is older than an already received RREP message), the node discards it directly; otherwise it calculates the energy levels of all neighboring nodes and determines their classes. Our idea is to classify nodes according to their energy levels. When the value of $w(u_j)$ of node $u_j$ is high and Equation *(7)* is used, this case is reduced to $e_{level}(u_j) \geq 1$.

Three classes of nodes are defined:

**Low**: The energy level of the node is below $\alpha$.
**Average**: The energy level of the node is between $\alpha$ and *1*.
**High**: The energy level of the node is greater or equal to *1*.

In the AOMDV protocol, the intermediate node establishes a reverse path by selecting the first neighboring node from the *route_list* field of the routing table (see Figure 1). In our protocol the intermediate node determines the class of each neighboring node stored in the table and forwards the RREP message to the neighboring node of the same class. If the classes of all neighboring nodes are different from the class of the intermediate node, the neighboring node with the lowest class above the class of the intermediate node is selected; if no higher class exists, the neighboring node with the highest class under the class of the intermediate node is selected. This process is repeated at every intermediate node until the source node. At the end of the process all discovered paths have same or near nodes classes, this results in homogeneous paths. The source node will have all paths energy information, this facilitates the paths selection for routing. To ensure link disjoint paths, an intermediate node marks each neighboring node selected by another reverse path; this property is ensured by the *marked_node* field of the routing table (see Figure 1); in addition our solution evaluates the energy level of each node, so the exhaustion of a node is considered and does not affect the links of paths. The trajectories of each RREP message may intersect at an intermediate node, but each takes a different reverse path to the source to ensure link-disjointness. When the node source receives the first RREP message for a destination, it waits a specified time, called *RREP_Wait_time*, before selecting the best path; this time is an additional delay for multipath routing protocols.

### 4.4 Data transmission phase

The choice of the best path between a source node *s* and destination node *d* depends on the energy levels of nodes. AOMDV transmits data packets using the first path from the list of available paths when multiple paths are established. In AOMR-LM protocol, paths are grouped by class (high, average, and low) according to the classes of nodes that build these paths. The paths of high class are selected first to forward the data packets. In this case, all packets going to the same destination follow the same path as long as there is no link failure in this path. Once the high class is empty, the algorithm selects any path of the average class. The same process is repeated with the low class. This process balances the nodes energy consumption which extends the network lifetime. One path is selected at a time for a data packet transmission between a source and a destination. If the selected path fails, the source node receives a RERR; in this case, a path is selected from the highest available class. Path discovery is initiated again when there are no more paths; this discovery process is described in Section 4.2.

### 4.5 Route maintenance

Route error detection in AOMR-LM is similar to route error detection in AOMDV. It is launched when a link fails between two nodes along a path from a source to a destination.
When a neighboring node does not respond to three successive HELLO messages sent by a node, the link is considered to have failed. If a node detects a failure of a link in an active path, it erases the route from its table and then sends an RERR message to the source node of the path. Each intermediate node returns this RERR message along the reverse path to the source node. When a source node receives an RERR message, it erases the path from its table and looks for an alternate path towards the destination node, if one is available; otherwise it initiates a path discovery process to resume the data transmission. An alternative path is selected as described in Section 4.3.

## 5 Performance Evaluation of AOMR-LM

In this section, we present simulation results to demonstrate the efficiency of our proposed protocol. First we present the metrics used for performance evaluation and then we analyze the values of coefficient $\alpha$ to select the most appropriate value for the rest of the simulations; this value allows us to classify nodes according to their energy levels. Finally we evaluate our protocol by comparing it with two protocols in the literature, namely AOMDV and ZD-AOMDV. This evaluation is accompanied with an analysis and discussion of results.

### 5.1 Performance parameters

We evaluate three key performance metrics. The network lifetime can be defined in three ways [38]: the time taken to exhaust the battery of the first network node, the time taken to exhaust the battery for *N* network nodes, and the time when the battery of the last network node is exhausted. We choose the second way. Energy consumption is the average of the energy consumed by nodes participating in packets transfer from the source node to the destination node during the whole simulation. End-to-end delay is the average transmission delay of data packets that are delivered successfully over the total duration of the simulation.

### 5.2 Analysis of the coefficient $\alpha$

In our protocol the choice of the path used by a source node to transmit data packets toward a destination node is based on the path class, which depends on the classes of nodes belonging to this path. In our protocol, the class definition of a node is mainly affected by the value of $\alpha$. In order to find appropriate values of $\alpha$, we apply Equations *(8)* and *(10)* to find the minimum values to define mainly two classes (low and average). The maximum value of $\alpha$ is determined by the equation (9), we have not studied this case, because in our work the third class (high) is determined by the value of $\beta$. We assume that if the number of nodes participating in the transfer of data is near to *K*, then the value of $T_{Net}$ approaches zero. Table 1 shows $\alpha_{min}$ values measured by varying *K* from 10 to 100 nodes.

**Table 1: Measuring $\alpha_{min}$ relative to the number K.**

| K | 10 | 20 | 30 | 40 | 50 | 60 |
|---|---|---|---|---|---|---|
| $\alpha_{min}$ | 0.155 | 0.289 | 0.396 | 0.499 | 0.574 | 0.629 |

| K | 70 | 80 | 90 | 100 |
|---|---|---|---|---|
| $\alpha_{min}$ | 0.672 | 0.707 | 0.734 | 0.757 |

In Figure 2 we can see that $\alpha_{min}$ increases with $K$, and tends toward zero as $K$ becomes small, therefore $\alpha_{min}$ is a function of $K$ when $T_{Net}$ tends to zero. For a network size of 190 nodes, the number of nodes used for transferring data packets from a source node to the destination node is between 30 and 40 (corresponding to $K$ of Table 1)[39].

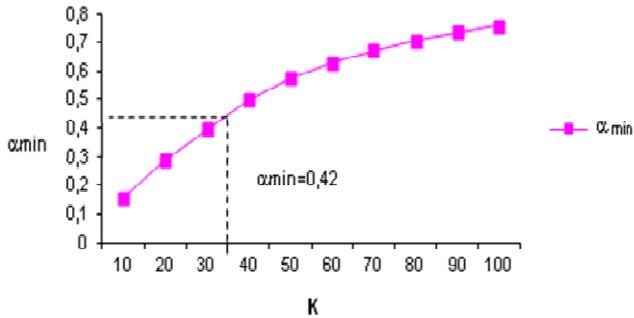

**Figure 2:** $\alpha_{min}$ **versus** $K$.

For the simulation of our protocol AOMR-LM, we chose 190 as the network number nodes; hence our choice of $\alpha_{min} = 0.42$.

### 5.3 Performance evaluation

We carried out simulations to determine the effectiveness of our protocol. The principal goal of these simulations is to analyze our protocol by comparing it with other protocols, mainly AOMDV [11] and ZD-AOMDV [34]. The values of simulation parameters are summarized in Table 2.

**Table 2: Simulation parameters.**

| Communication Model | Constant Bit Rate (CBR) |
|---|---|
| MAC type | IEEE 802.11 |
| Mobility model | Random Waypoint |
| Terrain range | 840 m × 840 m |
| Transmission range | 250 m |
| Number of mobile nodes | 30 to 190 |
| Data payload | 512 bytes |
| RREQ_Wait_Time | 1.0 s |
| RREP_Wait_Time | 1.0 s |

To evaluate AOMR-LM, we use the network simulator ns-2 [40]. Each simulation run has a duration of 300 seconds. During each simulation, constant bit rate (CBR) connections are generated; each of them produces four packets per second with a packet size of 512 bytes. The values of *RREQ_Wait_Time* and *RREP_Wait_Time* are set to 1.0 seconds, the same value as that used for the protocol AOMDV [11].

We vary the number of network nodes from 30 to 190 to obtain different scenarios in an 840 m × 840 m environment.

The Random Waypoint model is used to simulate node movement; each node moves with a speed randomly chosen from 0 to 5 m/s. The radio model uses characteristics similar to a commercial radio interface, Lucent's Wave LAN. Wave LAN [41] is a shared-media radio with a nominal bit-rate of 2 Mbit/s and a nominal radio range of 250 m, which is compatible with the IEEE 802.11 standard. Each simulation is carried out under a different number of network nodes and the performance metrics are obtained by averaging over 20 simulation runs from one source to one destination randomly selected. We assume that a node consumes 281.8 mW while receiving and 281.8 mW while transmitting [42]. It was shown in [43] that no real node energy optimization can be achieved in the presence of overheating or in idle state. For this reason, the energy consumption during idle or overheating time is not considered in this model. In our simulations, we initialized the energies of the nodes randomly between 10 and 60 Joules (uniform distribution), which corresponds to the average capacity of a battery.

We evaluate the performance of AOMR-LM by comparing it with the AOMDV and ZD-AOMDV routing protocols. The network lifetime metric is shown in Figure 3 with the number of network nodes equal to 190. The network lifetime of AOMR-LM is longer than those of AOMDV and ZD-AOMDV. Our protocol exhausts fewer nodes compared to ZD-AOMDV and AOMDV protocols, which increases the lifetime of the network.

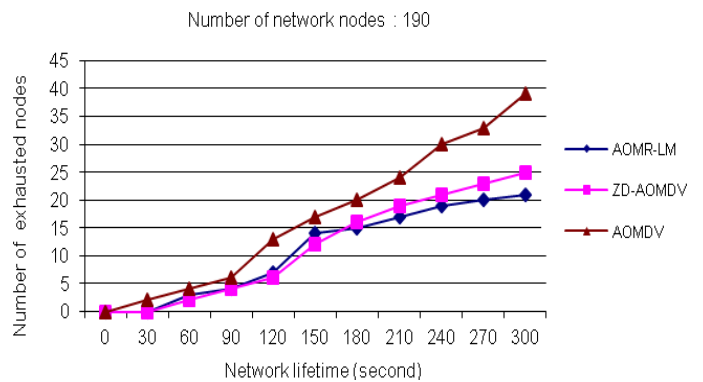

**Figure 3: Network lifetime versus number of exhausted nodes.**

Thus AOMR-LM balances the energy among all the nodes and prolongs the individual node lifetime and hence the entire network lifetime.

Figure 4 shows the energy consumed in different scenarios by the AOMR-LM, ZD-AOMDV, and AOMDV protocols. AOMR-LM does not perform too well at the beginning of the simulation, but it improves later. Initially, it is not better than any other protocol, because initially the majority of data packets are not yet transmitted, so the total energy of sending and receiving packets is not important.

But in the last stage, as time increases, there is some imbalance of energy that comes into play and then the impact of our algorithm comes into play. We can see that the energy consumed in AOMR-LM is less than those consumed by ZD-AOMDV or AOMDV.

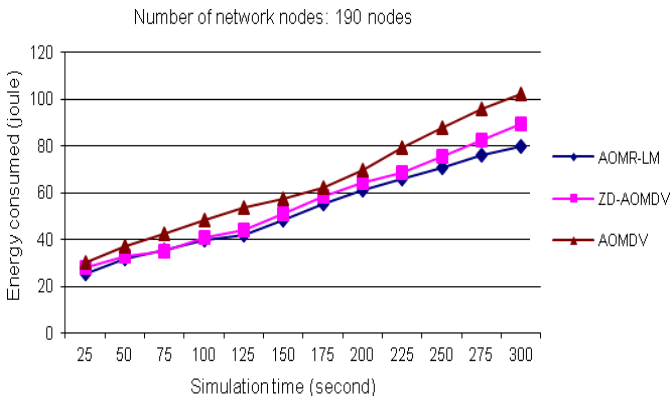

**Figure 4: Energy consumed versus time.**

AOMR-LM consumes less energy than ZD-AOMDV or AOMDV, firstly because AOMR-LM is able to balance the energy between paths. Thus energy is balanced out across the network, reducing uneven energy consumption. Secondly, AOMR-LM is able to avoid nodes with low energy in the construction of the multipath. This means that paths with higher energy are identified and selected for transmission.

Figure 5 shows the average end-to-end delay of the compared protocols. The average end-to-end delay for all tested protocols increases when increasing the network size, but the average end-to-end delay of AOMR-LM is lower than those of ZD-AOMDV and AOMDV, for different nodes speeds, see Figures (a) and (b).

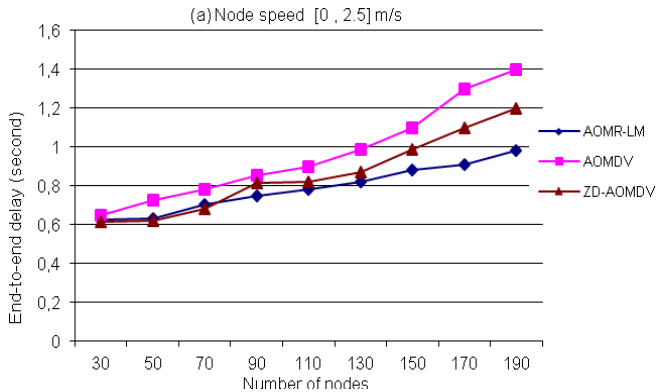

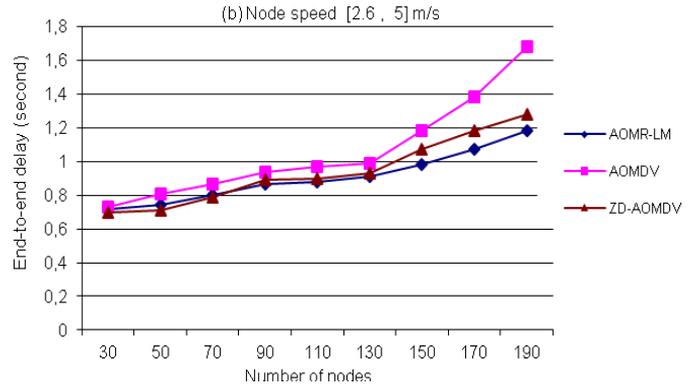

**Figure.5: End-to-end delay versus number of nodes.**

When the number of nodes is less than 70 Figure 5 (a) and (b), the protocol ZD-AOMDV protocol has a delay equal or slightly better than our protocol, due to the simultaneous use of paths. Once the size of the network increases, our protocol produces a better delay, this for any nodes speed. The reason is that our AOMR-LM protocol favors nodes having a high energy level and prevents the critical nodes from participating in the data packet transmission. This produces fewer broken links and greatly reduces the end-to-end delay.

## 6 Conclusion

In this article we have provided a solution to the problems of routing in an ad hoc network. Mobile ad hoc networks are characterized by their lack of infrastructure and their dynamicity: link failures and route breaks occur frequently. Moreover, the frequent changes of topology exhaust the batteries of the nodes, which decreases the network performance. A new multipath routing protocol, AOMR-LM, has been proposed in this paper, performing energy aware routing in mobile ad hoc networks. We has shown that AOMR-LM conserves the residual energy of nodes and balances the consumed energy over multiple paths. AOMR-LM routing protocol is an extension of the existing multipath routing protocol AOMDV. It uses an energy-aware mechanism, which exploits the residual energy of nodes to select and classify the paths according to the energy level of their nodes. This concept extends the network lifetime and improves energy consumption when compared with other solutions known in the literature. The coefficient $\alpha$ is analyzed in order to find appropriate values, which are required to define the class of a node during the reply-forwarding process and to preserve the node residual energy. Comparing the performance of AOMR-LM with those of the AOMDV and ZD-AOMDV protocols, AOMR-LM is able to balance the energy consumed. It increases the lifetime, consumes less energy and has a lower average end-to-end delay than the other simulated protocols, because paths are computed depending on the energy level of their nodes, and the one of the best paths is selected. In the future we plan to study the cooperation of our routing protocol with a MAC layer power-control technique to see how they can cooperate to decrease the energy consumption of ad hoc networks.